\documentclass[12pt]{article}
\newwrite\ffile\global\newcount\figno \global\figno=1

\def\writedef#1{}

\input epsf
\def\figin{\epsfcheck\figin}\def\figins{\epsfcheck\figins}
\def\epsfcheck{\ifx\epsfbox\UnDeFiNeD
\message{(NO epsf.tex, FIGURES WILL BE IGNORED)}
\gdef\figin##1{\vskip2in}\gdef\figins##1{\hskip.5in}
\else\message{(FIGURES WILL BE INCLUDED)}%
\gdef\figin##1{##1}\gdef\figins##1{##1}\fi}

\def\figinsert{}
\def\ifig#1#2#3{\xdef#1{fig.~\the\figno}
\writedef{#1\leftbracket fig.\noexpand~\the\figno}%
\figinsert\figin{\centerline{#3}}\medskip\centerline{\vbox{\baselineskip12pt
\advance\hsize by -1truein\center\footnotesize{ Fig.~\the\figno.} #2}}
\bigskip\endinsert\global\advance\figno by1}
\def\endinsert{}

\begin{document}
\baselineskip 18pt
\newcommand{\Tr}{\mbox{Tr\,}}
\newcommand{\beq}{\begin{equation}}
\newcommand{\eeq}{\end{equation}}
\newcommand{\bea}{\begin{eqnarray}}
\newcommand{\eea}[1]{\label{#1}\end{eqnarray}}
\renewcommand{\Re}{\mbox{Re}\,}
\renewcommand{\Im}{\mbox{Im}\,}
\begin{titlepage}

\begin{picture}(0,0)(0,0)
\put(350,0){SHEP-02-14}
\end{picture}
 
\begin{center}
\hfill
\vskip .4in
{\large\bf A Non-supersymmetric Deformation of the AdS/CFT Correspondence}
\end{center}
\vskip .4in
\begin{center}
{\large James Babington, David E. Crooks and Nick Evans}
\footnotetext{e-mail: dc@hep.phys.soton.ac.uk, jrb4@hep.phys.soton.ac.uk, 
n.evans@hep.phys.soton.ac.uk }
\vskip .1in
{\em Department of Physics, Southampton University, Southampton,
S017 1BJ, UK}

\end{center}
\vskip .4in
\begin{center} {\bf ABSTRACT} \end{center}
\begin{quotation}
\noindent We deform the AdS/CFT Correspondence by the inclusion of a
non-supersymmetric scalar mass operator. We discuss the 
behaviour of the dual 5
dimensional supergravity field then lift the full solution to 10 dimensions.
Brane probing the resulting background reveals a potential consistent 
with the operator we wished to insert. 

\end{quotation}
\vfill
\end{titlepage}
\eject
\noindent

\section{Introduction}

A corner stone of the AdS/CFT Correspondence \cite{mald, kleb,ed}
is that the supergravity
fields in $AdS_5 \times S^5$ act as source terms for operators in the
${\cal N}=4$ gauge theory dual. It is essential that there is a one to 
one mapping between the supergravity fields and these operators. A
corollary of this relationship is that we can study any deformation
of the ${\cal N}=4$ gauge theory since we can introduce all 
possible operators. 

An industry has grown in attempting to generate gravity duals of all
interesting deformations of the AdS/CFT Correspondence 
\cite{gppz1,gppz2,dz,freed1,gub2} and understand how
different field theory phenomena are encoded gravitationally. Much
of the early work in this respect has concentrated on
supersymmetric deformations such as the ${\cal N}=4$ theory on
moduli space \cite{freed2}, ${\cal N}=1$ Leigh Strassler 
theory \cite{pw2,ls}, the ${\cal N}=2^*$ \cite{pw,bs2,ep,ejp,bpp}
and ${\cal N}=1^*$ theories \cite{gppz3,ps,pw2}. Allowing
non-trivial solutions for the supergravity fields is relatively
straightforward at the 5d supergravity level. Connecting the resulting
backgrounds to the field theory has proven difficult with more success
being obtained in the cases where the solutions have been lifted to
10d. The technology to lift solutions to 10d \cite{pilch,freed2,pw,pw2}
has been developed but
in more complicated cases such as the ${\cal N}=1^*$ theory \cite{pw2}
the complexity
can grow to prohibit a full solution being found. The major benefit
of a 10d solution is that brane probing \cite{mald,ejp,bpp,ls,more,martelli,
abc} can be used to directly convert
the space-time background into a U(1) gauge theory on the surface of the
probe. In this way the gauge coupling, operator parametrization and
potentials can be successfully investigated. It should be noted that, since
the ${\cal N}=4$ gauge theory is strongly coupled in the ultra-violet,
fields can never be decoupled from the strong interactions by making them
massive in these models. 

More recently interest has turned to non-supersymmetric deformations
of gauge/gravity duals \cite{dz,nsgub,of,epz,nsgub2}. 
It is interesting to see whether the dualities
continue to make sense without supersymmetry and there is also the 
potential to investigate new phenomena not present in supersymmetry.
Apart from an early paper on non-supersymmetric deformations in 5d \cite{dz},
recent attention \cite{nsgub,of,epz,nsgub2}
has focused on deformations of more involved ${\cal N}=1$
supersymmetric constructions such as the Maldecena Nunez \cite{mn} and the
Klebanov Strassler \cite{ks} backgrounds. These theories have discrete 
vacua and hence supersymmetry breaking perturbations will not result in
an unstable background. In this paper we return to deforming the orginal
AdS/CFT correspondence. We will in fact introduce a mass term of the form
$(\phi_1^2 + \phi_2^2 + \phi_3^2 + \phi_4^2 - 2 \phi_5^2 -2 \phi_6^2)$
which is naively unbounded. Our interest is in developing the technology
to find and lift these solutions to 10d so we will not be so concerned by
the runaway behaviour (although the 10d solution we provide correctly
reproduces the expected behaviour). One might hope that there would
be such backgrounds that
are really stable since an SO(6)$_R$ singlet scalar
mass term is not visible in the supergravity solution as it is not
in a short multiplet. It's presence could stabilize the solution.
Note that the supersymmetric deformations \cite{gppz2,pw,pw2}
already mentioned require this operator to be present. In fact our brane 
probe potential reveals the operator not to be present in our 10d lifts.
Our solution is also of interest since it is probably the simplest
example of a non-supersymmetric deformation; only the metric and
four potential fields are non-zero.

In the next section we will discuss the introduction of our deformation
at the 5d supergravity level. In section 3 we then lift the full solution
to 10d, although one function in the four form is only found numerically. 
In section 4 we brane probe the background with a D3 brane
and show that 
asymptotically the background indeed includes the operator we hoped 
to introduce showing the consistency of the techniques. Finally we plot
the potential seen by the probe for the full solution.

\section{Deformations in 5d Supergravity}

According to the standard AdS/CFT Correspondence map \cite{kleb,ed}
each supergravity
field plays the role of a source in the dual field theory. The simplist
possibility is to consider non-trivial dynamics for a scalar field in
the 5d supergravity theory. We only allow the scalar to vary in the radial 
direction in AdS with the usual interpretation that this corresponds to 
renormalization group running of the source. As is standard in the literature 
\cite{gppz1,dz} we look for solutions where the metric is described by

\beq
ds^2 = e^{2 A(r)}dx^\mu dx_\mu + dr^2
\eeq
where $\mu=0..3$ and $r$ is the radial direction in AdS$_5$. The scalar
field has a lagrangian

\beq
{\cal L} = {1 \over 2} (\partial \lambda)^2 + V(\lambda)
\eeq

There are two independent, non-zero, elements of the Einstein tensor ($G_{00}$
and $G_{rr}$) giving two equations of motion plus there is the 
usual equation of motion for the scalar field \cite{gppz1}

\beq \label{e1}
\lambda^{''} + 4 A^{'} \lambda^{'} = {\partial V \over \partial \lambda}
\eeq

\beq \label{e2}
6 A^{'2} = \lambda^{'2} - 2 V
\eeq

\beq \label{e3}
-3 A^{''} - 6 A^{'2} = \lambda^{'2} + 2 V
\eeq

In fact only two of these equations are independent but it will be useful 
to keep track of all of them. 

In the large $r$ limit, where the solution will return to AdS$_5$ at first
order and $\lambda \rightarrow 0$ and $V \rightarrow m^2 \lambda^2$,
only the first equation survives with solution

\beq
\lambda = {\cal A} e^{-\Delta r} + {\cal B} e^{-(4 - \Delta) r}
\eeq
with

\beq
m^2 = \Delta(\Delta-4)
\eeq
${\cal A}$ is interpreted as a source for an operator and ${\cal B}$
as the vev of that operator since $e^r$ has conformal dimension 1.

If the solution retains some supersymmetry then the potential can be written 
in terms of a superpotential \cite{freed1}

\beq
V = { 1 \over 8} \left| { \partial W \over \partial \lambda} \right|^2 - 
{1 \over 3} |W|^2
\eeq
and the second order equations reduce to first order

\beq \label{susyeom}
\lambda^{'} = { 1 \over 2} {\partial W \over \partial \lambda}, \hspace{1cm}
A^{'} = - {1 \over 3} W
\eeq

\subsection{A Scalar Operator}

Let us now make a particular choice for the scalar field we will consider.
We take a scalar from the multiplet in the 20 of SO(6). 
These operators have been identified \cite{ed} as playing the role 
of source and vev for the scalar operator $tr \phi_i \phi_j$ in the field 
theory. In particular we will chose the scalar corresponding to the 
operator

\beq \label{op}
{\cal O} = \phi_1^2 + \phi_2^2 + \phi_3^2 + \phi_4^2 - 2 \phi_5^2 - 2 \phi_6^2
\eeq
This scalar has been studied in the literature \cite{freed2,more} 
already in its role of
describing an ${\cal N}=4$ preserving scalar vev and as a mixture of a
mass term and a vev in the ${\cal N}=2^*$ gauge theory \cite{pw,bs2}. 
The potential
for the scalar, which we will write as $\rho = e^{\lambda/\sqrt{6}}$ 
is given by

\beq \label{pot}
V = - {1 \over \rho^4} - 2 \rho^2
\eeq
and the three equations of motion become

\beq
{ \rho^{''} \over \rho} - \left( {\rho^{'} \over \rho}\right)^2 
+ 4 {\rho^{'} \over \rho} A^{'} = {\rho \over 6} { \partial V \over \partial \rho}
\eeq

\beq
6 A^{'2} - 6 \left( {\rho^{'} \over \rho}\right)^2 = - 2 V
\eeq

\beq
A^{''} = -4 \left( {\rho^{'} \over \rho}\right)^2
\eeq
The last of these is the sum of (\ref{e2}) and (\ref{e3}). 
The asymptotic ($r \rightarrow 
\infty$) solutions take the form
 
\beq
\lambda = {\cal A} e^{-2 r} + {\cal B}r e^{-2 r}
\eeq
with ${\cal A}$ the scalar vev and ${\cal B}$ a mass term for the 
operator ${\cal O}$.

In the special case where only the first part of the solution
is present the deformation preserves ${\cal N}=4$ supersymmetry. The
superpotential is

\beq
W = - {1 \over \rho^2} - {1 \over 2} \rho^4
\eeq
and the
second order equations reduce to the first order equations

\beq \label{susy}
{\partial \rho \over \partial r} = {1 \over 3} \left({ 1 \over \rho}-
\rho^5\right), \hspace{1cm} {\partial A \over \partial r} = 
{2 \over 3} \left({ 1 \over \rho^2}+ {1 \over 2}
\rho^4\right)
\eeq
with solution \cite{freed2}
\begin{equation} \label{Atorho}
e^{2A} = l^2 {\rho^4 \over \rho^6-1}
\end{equation}
with $l^2$ a constant of integration.

\subsection{Non-supersymmetric First Order Equations}

In \cite{ver} it was pointed out that using Hamilton Jacobi theory the second
order equations could be replaced by a system of first order equations.
They further stated that a ``superpotential'', $W$, could be found
which resulted in the equations (\ref{susyeom}) even for the non supersymmetric
solution with only ${\cal B}$ switched on. A similar result was obtained in 
\cite{town,cvet} but as a requirement for the RG flow solution to be
stable. Further analysis along these lines can be found in \cite{dario,dario2}.
Reducing the equations to first order would be very 
helpful, but the system we discuss here can not be.

Consider the UV of the theory where, expanding (\ref{pot})

\beq
V = -3 - 2 \lambda^2 + \sqrt{8 \over 27} \lambda^3 + ...
\eeq
we can attempt to find a superpotential $W$ that reproduces this 
potential via the trial form

\beq
W = a + b \lambda^2 + c \lambda^3 + ...
\eeq
Working to quadratic order one finds

\beq
a=-3, \hspace{1cm} b = - 2
\eeq
The solution for $b$ comes from a quadratic equation with degenerate
roots hinting at the two forms of the solution. 
However, it is then easy to show that at higher orders there is a unique
series (eg $c= \sqrt{2/27}$) and it is simply the
supersymmetric solution. We have therefore not been able
to find a superpotential that describes the non-supersymmetric solution and
are forced to numerically solve the second order equations. Of course
our geometry is intrinsically unstable since we have introduced an unbounded
operator in the field theory. Apparently the stability of the flow is 
essential for the system to reduce to first order.

\begin{center}
\hskip-10pt{\lower15pt\hbox{
\epsfysize=2.5 truein \epsfbox{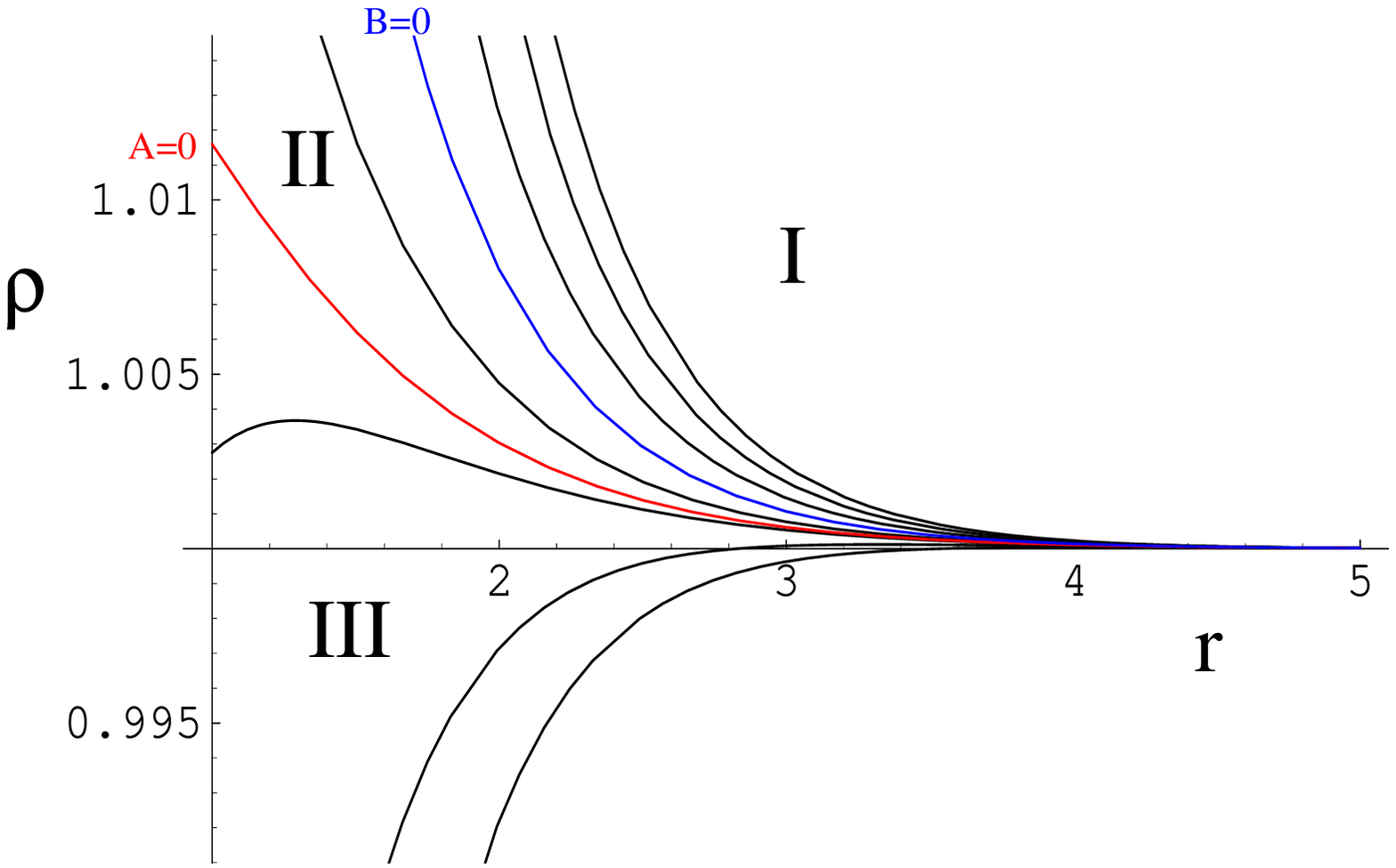}}}\medskip

Figure 1: Plots of $\rho$ vs $r$ for a variety of initial conditions
on $\rho^{'}$. The vev only initial condition solution is marked
with ${\cal B}=0$ and mass only initial condition with ${\cal A}=0$. 
The marked regions are explained in (\ref{region}).
\end{center}

\subsection{Numerical Solutions}

The second order equations of motion are easily solved.
In figure 1 we show the numerical behaviour of $\rho$. For this plot we fix
$\rho(r=\Lambda_{UV})$ and vary the derivative. 
The purely vev supersymmetric solution 
(${\cal B}=0$) and purely masslike case (${\cal A}=0$) are labelled.
The three regions (bounded by the ${\cal A}=0$ and ${\cal B}=0$ curves)
correspond to 

\beq \label{region}
\begin{array}{ccc}
& {\cal A} & {\cal B} \\
&& \\
I & +ve & -ve\\
II & +ve & +ve\\
III & -ve & +ve
\end{array}
\eeq

In all these cases the function $A(r)$ only deviate from each other
and $A(r)\sim r$ by a small amount so a plot is unrevealing. 
Note that most of these solutions become
singular before $r=0$. When lifted to 10d this singular point
is expected \cite{freed2} to correspond to the position of the D3 brane sources
in the background. For most of these solutions there is a scalar vev 
and so the D3 branes are expected to have moved away from the origin.
The mass only solution (${\cal A}=0$) on the other hand can be extended
to $r=0$ which is consistent with the D3 branes being pinned at the origin.

It has proven difficult to extract aspects of the field theory from the 5d
supergravity backgrounds. More success has been had at the 10d level where 
techniques such as brane probing can be used to connect to the field
theory. We shall therefore move to discussing the lift of these solutions
to 10d in the next section.

\section{The 10d Background}

To lift the 5d solution to 10d requires the procedure outlined in \cite{pilch}.
Finding the metric is complicated but we will be able to short
cut the process since the lift of the 5d solution where the ${\cal N}=4$
theory is on moduli space has already been written down. 
In particular
the solution where our scalar corresponds to a vev has been studied in 
\cite{freed2,more}
(it is also the limit of the metrics in \cite{pw,bs2,pw2} 
with some of the fields
switched off). That solution is given by

\begin{equation} \label{4met}
ds^2=\frac{X^{1/2}}{\rho}e^{2A(r)} dx_{//}^2 +
\frac{X^{1/2}}{\rho}
\left(dr^2+\frac{L^2}{\rho^2}\left[d\theta^2+\frac{\sin^2\theta
}{X
}d\phi^2+\frac{\rho^6\cos^2\theta}{X}d\Omega^2_3\right] \right),
\end{equation}
where $d\Omega^2_3$ is the metric on a 3-sphere and
\begin{equation}
X \equiv \cos^2\theta+\rho^6\sin^2\theta
\end{equation}
For consistency there must also be a non-zero $C_4$ potential of the form
\begin{equation} \label{4C4}
C_4 = \frac{e^{4A}X}{g_s \rho^2} dx^0 \wedge dx^1 \wedge dx^2 \wedge dx^3
\end{equation}
Note that the solution has the same SO(2) $\times$ SO(4) 
symmetry as our operator (\ref{op}). 

Clearly the lift of the full solution of the second order equations has
this as a limit. In fact the procedure for finding the form of the metric
does not depend on the supersymmetric solution and we may take it over
directly to our case. The $C_4$ potential though will change since the
supersymmetric first order equations of motion were used in its derivation
\cite{freed2,pw}.

In fact the 10d supergravity equations of motion we must concern 
ourselves with are relatively few \cite{pw} since only the metric and $C_4$ 
are non-zero. There are the Einstein equations

\beq
R_{MN} = T_{MN} = {1 \over 6} F^{PQRS}_N F_{PQRSM}
\eeq
and

\beq \label{bi}
F_{(5)}= ^*F_{(5)}, \hspace{1cm} d F_{(5)} = 0
\eeq
The self duality condition can be imposed by using the ansatz

\beq
F_{(5)} = {\cal F} + ^*{\cal F}, \hspace{1cm} 
{\cal F} = dx^0 \wedge dx^1 \wedge dx^2 \wedge dx^3 \wedge d w
\eeq
where $w(r,\theta)$ is an arbitrary function. 

There are three independent non-zero elements of $R_{MN}$ which factorize
into the useful equations

\beq \label{dwdt}
R^0_0 - R^r_r = {1 \over 2} 
g^{00} g^{11} g^{22} g^{33} g^{rr} \left( { \partial w
\over \partial r}\right)^2
\eeq

\beq
R^0_0 + R^r_r = {1 \over 2} g^{00} g^{11} g^{22} g^{33} g^{\theta \theta} 
\left( { \partial w \over \partial \theta}\right)^2
\eeq

\beq
R^r_\theta = {1 \over 2} g^{11} g^{22} g^{33} g^{44} g^{rr} 
\left( { \partial w \over \partial \theta} { \partial w
\over \partial r}\right)
\eeq

The right hand side of these equations are straightforward but 
laborious to explicitly calculate. The resulting output is lengthy but
can be simplified by using the second order
equations of motion to eliminate $\rho^{''}, A^{''}$ and $A^{'2}$. The 
resulting background will therefore reproduce the full second order
equations of motion. We find

\beq
R^0_0 - R^r_r = - {18 \sin^2 \theta \cos^2 \theta \rho^5 \rho^{'2}
\over X^{5/2}}
\eeq

\beq
R^0_0 + R^r_r =- {(2 \cos^2 \theta - (\cos 2 \theta - 3) \rho^6)^2
\over 2 \rho^3 X^{5/2}}
\eeq

\beq
R^r_\theta = - { 3 \sin^2 \theta \rho^{'} ( ( \cos 2 \theta - 3) \rho^6 - 2
\cos^2 \theta) \over 2 X^{5/2}}
\eeq

(\ref{dwdt}) thus reduces to

\beq
\left( {\partial w \over \partial \theta} \right) = {6 e^{4 A} \cos \theta
\sin \theta \rho^{'} \over \rho}
\eeq
which can be directly integrated and $w$ put in the form

\beq
w(r, \theta) = {e^{4 A} \over
\rho^2} - { 3 \sin^2 \theta \rho^{'} e^{4 A} \over \rho}- e^{4A} F(r)
\eeq
where $F(r)$ is as yet undetermined. 

Note that the supersymmetric limit corresponds to $F(r)=0$ and $\rho^{'}$
replaced using the supersymmetric first order equation of motion (\ref{susy}). 
We should not
be surprised that derivatives of $\rho$ enter directly into the solution 
since introducing a mass term corresponds explicitly to introducing an extra
degree of freedom via precisely this derivative. 

$F$ can then be found using either of the other two equations (the third
equation providing a check on the consistency of the solution). It is the
solution of

\beq \label{f}
- 2 - 2 \rho^6 = - 4 \rho^2 A^{'} + 4 \rho^4 F A^{'} + \rho^4 F^{'} + 2 \rho
\rho^{'}
\eeq
We have not been able to solve this equation explicitly but in the UV limit
the solution takes the form

\beq \label{bcf}
F = {1 \over 3} \left( {1 \over \rho} - \rho^5 \right) - \rho^{'} + ....
\eeq
which clearly vanishes in the supersymmetric limit given (\ref{susy}).
For a general numerical solution of the second order equations of motion
we can set 
the boundary conditions on $F$ using this asymptotic form and hence find 
$F$ numerically for all $r$. 

The solution then faces its strongest test since $F_{(5)}$ must also
satisfy its bianchi identity (\ref{bi}). At first sight this appears to be a challenge; since $w$ contains a derivative
of $\rho$ the bianchi identity is a third order equation. 

The Bianchi identity (\ref{bi}) is (since $d {\cal F} = 0$)
\beq
dF=\left(\partial_{r}[\sqrt{g} g^{00}g^{11}g^{22}g^{33}g^{rr}\partial_{r}\omega] +\partial_{\theta}[\sqrt{g}
g^{00}g^{11}g^{22}g^{33}g^{\theta \theta}\partial_{\theta}\omega] \right)
d\Omega_{5}\wedge dr=0
\eeq

Hence we must check that

\beq
\partial_{r}[X^2 e^{-4A} \rho^{4} \sin \theta \cos^3 \theta \sqrt{ \det S_{3}} \partial_{r}\omega]+\partial_{\theta}[X^2
e^{-4A} \rho^{6} \sin \theta \cos^3 \theta \sqrt{ \det S_{3}} \partial_{\theta}\omega]=0
\eeq

In fact explicit computation,
using the second order equations
of motion and (\ref{f}), shows that this third order equation 
is satisfied and the solution survives.

Given the complete numerical
10d lift of our non-supersymmetric solutions we can study
the background for signals that it correctly encodes the field theory
dynamics.

\begin{center}
\hskip-10pt{\lower15pt\hbox{
\epsfysize=2.5 truein \epsfbox{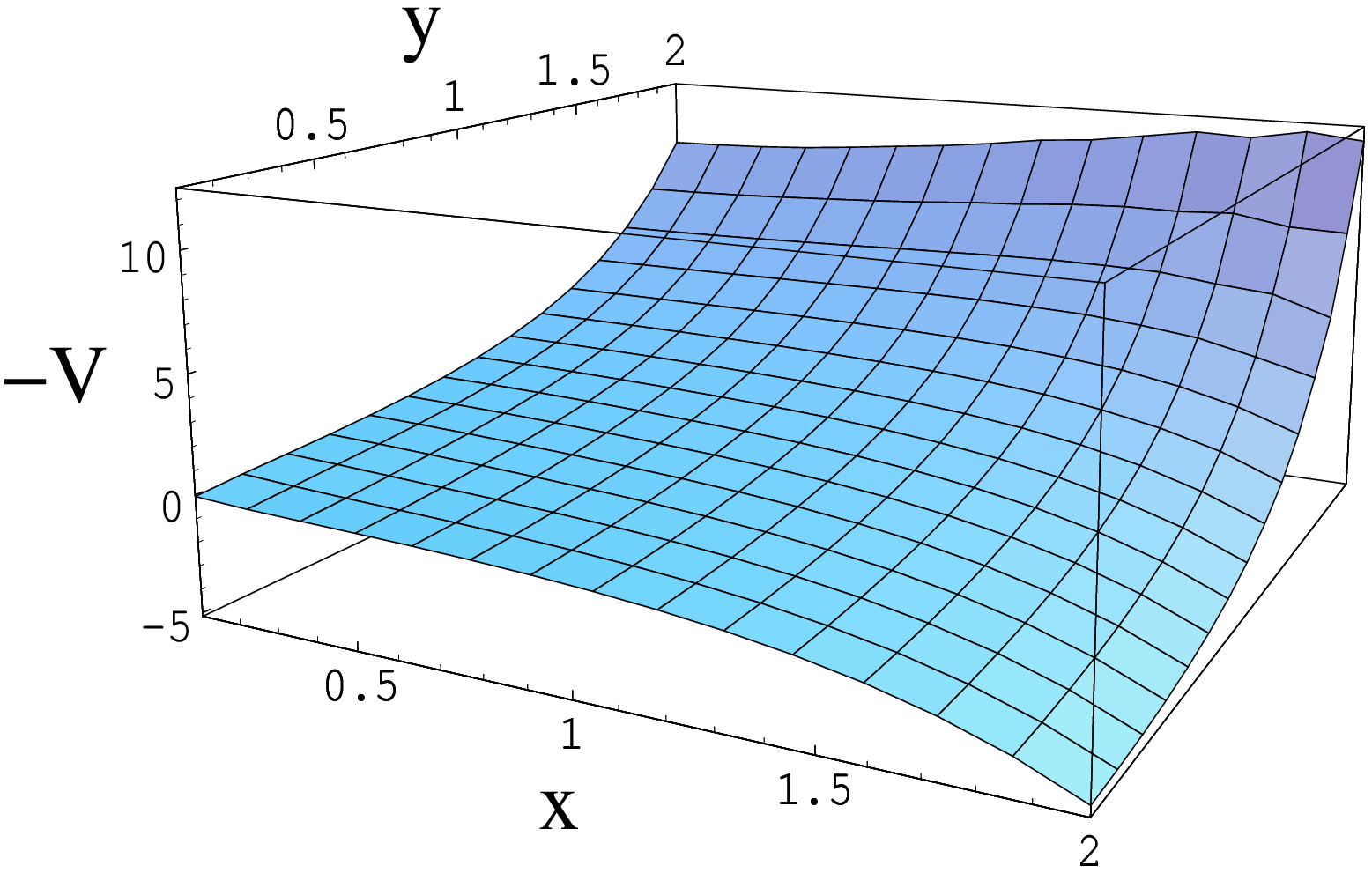}}}\medskip

Figure 2: The probe potential plotted over the $r-\theta$ plane 
for the mass only case (${\cal A}=0$).
\end{center}

\section{Brane Probe Potential}

The most succesful technique for connecting backgrounds and their dual
field theories has been brane probing \cite{mald,ejp,bpp,ls,more,martelli,
abc}
which converts the background
to the U(1) theory on the probe's surface. We thus substitute the background 
into the Born-Infeld action

\begin{equation} \label{BI}
S_{probe}=-\tau_3\int_{\mathcal{M}_4}d^4x
\det[G^{(E)}_{ab} + 2 \pi \alpha' e^{- \Phi/2} F_{ab}]^{1/2} 
+ \mu_3\int_{\mathcal{M}_4} C_4,
\end{equation}
The resulting scalar potential is given by

\beq
V_{probe} = - e^{4A} \left[ {X \over \rho^2} + {3 \sin^2 \theta \rho^{'} \over 
\rho} - {1 \over \rho^2} + F \right]
\eeq

It is illuminating to evaluate this potential at leading order in the 
UV with

\beq
\rho = 1 + v e^{-2 r} + m^2 r e^{-2r} + ...
\eeq
We find

\beq
V = m^2 e^{2 r} ( 2 - 6 \sin^2 \theta) + ...
\eeq

The scalar vev vanishes from the potential at this order consistent 
with the existence of the ${\cal N}=4$ moduli space. The mass term
reproduces precisely the mass operator we expected in (\ref{op})
remembering that $e^r$ plays the role of a scalar field. We conclude
that the 10d background shows all the correct behaviour to be
dual to the non-supersymmetric gauge theory with scalar masses. 

Finally we numerically find the full solution for $A(r)$, $\rho(r)$ and
$F(r)$ for the mass only boundary conditions 
(${\cal A}=0$) using (\ref{e1},\ref{e2},
\ref{e3}) and (\ref{f},\ref{bcf}). We then
plot the full solution
for the probe potential in the $r-\theta$ plane
for the mass only solution (${\cal A}=0$) in figure 2. The plot 
fits well with the claim that the mass operator (\ref{op}) is present. 
The supersymmetric solutions (${\cal B}=0$) give a flat probe potential.
Other non-supersymmetric solutions reproduce the form of figure 2 upto a 
sign change dependent on the sign of ${\cal B}$. 

We conclude that we have successfully found the 10d gravity dual of this
simple non-supersymmetric deformation of the AdS/CFT Correspondence.
\vspace{0.5cm}

\noindent {\bf Acknowledgements:} NE is grateful to PPARC for the support
of an Advanced Fellowship. JRB and DC are grateful to PPARC for the support
of studentships.

\end{document}